\documentclass{article}

\usepackage{spconf}
\usepackage{amsmath,bm,cite,graphicx,amssymb,pifont}
\usepackage{makecell,arydshln,booktabs}

\usepackage{balance}

\title{An Investigation of Monotonic Transducers\\for Large-Scale Automatic Speech Recognition} 
\name{Niko Moritz, Frank Seide, Duc Le, Jay Mahadeokar, Christian Fuegen} 

\address{Meta AI}

\begin{document}

\maketitle
\begin{abstract} 

The two most popular loss functions for streaming end-to-end automatic speech recognition (ASR) are RNN-Transducer (RNN-T) and connectionist temporal classification (CTC). Between these two loss types we can classify the monotonic RNN-T (MonoRNN-T) and the recently proposed CTC-like Transducer (CTC-T).
Monotonic transducers have a few advantages. 
First, RNN-T can suffer from runaway hallucination, where a model keeps emitting non-blank symbols without advancing in time. 
Secondly, monotonic transducers consume exactly one model score per time step and are therefore more compatible with traditional FST-based ASR decoders. However, the MonoRNN-T so far has been found to have worse accuracy than RNN-T. It does not have to be that way: By regularizing the training via joint LAS training or parameter initialization from RNN-T, both MonoRNN-T and CTC-T perform as well or better than RNN-T. This is demonstrated for LibriSpeech and for a large-scale in-house data set.

\end{abstract}


\vspace{-2mm}
\section{Introduction}
\vspace{-1mm}

In recent years, the recurrent neural network-transducer (RNN-T) \cite{Graves12} has become one of the most important training criterion in automatic speech recognition (ASR) \cite{gulati2020conformer,Li2020DevelopingRM,Li2020OnTC}. Note that we loosely refer to RNN-T as the loss function in this work, which does not imply that any kind of RNN must be used for the model architecture.
Other popular and related training criteria are the connectionist temporal classifications (CTC) loss \cite{GravesFGS06}, the cross-entropy (CE) loss for encoder-decoder, traditional hybrid DNN/HMM ASR systems, and discriminative objectives such as MMI \cite{Vesel2013SequencediscriminativeTO,povey_thesis,povey16_interspeech}.
RNN-T is preferred over other ASR paradigms for three reasons:
(1) RNN-T models demonstrate better or competitive performance even without the use of an external language model \cite{ImprovRNNT2019};
(2) streaming recognition can naturally be accomplished with RNN-T based systems \cite{Rao2017ExploringAD,Sainath2020ASO}; and
(3) inference for RNN-T based ASR can be very efficient, making on-device large-vocabulary ASR possible \cite{He2019e2eASRmobile,MahaveerSM19}.

One special property of RNN-T is that it does not enforce strictly monotonic alignments between input frames and output labels. This can result in a particular behaviour where during inference the model can decide to output blanks (nothing) for a long period of time, followed by an emission of multiple labels in a single time step \cite{MonoRNNT19}. This is contrary to traditional ASR systems and to the strictly monotonic alignment of speech sounds.
To mitigate this, the monotonic RNN-T (MonoRNN-T) loss has been proposed \cite{MonoRNNT19}. It modifies the RNN-T loss as to restrict the summation of probabilities in the forward-backward pass to strictly monotonic alignments.
However, thus far it is found to perform less accurate than RNN-T \cite{JayAlignRest2021,MonoRNNT19,moritz2021sequence,zeyer2020new}, which presumably hinders it from a wider impact.

Recently, the graph temporal classification-transducer (GTC-T) loss function has been proposed \cite{moritz2021sequence}. GTC-T generalizes the RNN-T loss to arbitrary monotonic alignment topologies by specifying the label transition rules as a graph, instead of as hard-coded rules as part of the definition.
GTC-T has enabled the exploration of alternative alignments, including a transducer model with CTC-like transition rules named CTC-T.
Other graph-based full-sum loss functions, such as GTC \cite{Moritz2021SemiSupervisedSR} and GTN \cite{hannun2020dwfst}, are closely related to GTC-T. However, those cannot be used for training transducer-based ASR models that have an internal decoder (predictor) for modeling conditional dependencies of output labels.

In \cite{moritz2021sequence}, promising results are demonstrated for monotonic transducers with a CTC-like lattice as well as for MonoRNN-T, especially when decoding with an external language model (LM). 
However, experiments were conducted only on academic data sets (LibriSpeech and HKUST \cite{librispeech,hkust}), leaving it unclear whether monotonic transducers can compete with RNN-T on larger real-life data sets.
\cite{moritz2021sequence} also showed the importance of initializing model parameters for monotonic transducer training in order to obtain robust results.

In this work, we are investigating GTC-T with a CTC-like and a MonoRNN-T graph using a large-scale data set of 145K hours. Different training strategies are proposed and evaluated. Furthmore, we discuss the reasons why better training strategies are required for monotonic transducers in Section~\ref{sec:discussion}.

\section{Transducer-based ASR}

\begin{figure*}[t]
  \centering
  \centerline{\includegraphics[width=0.90\linewidth]{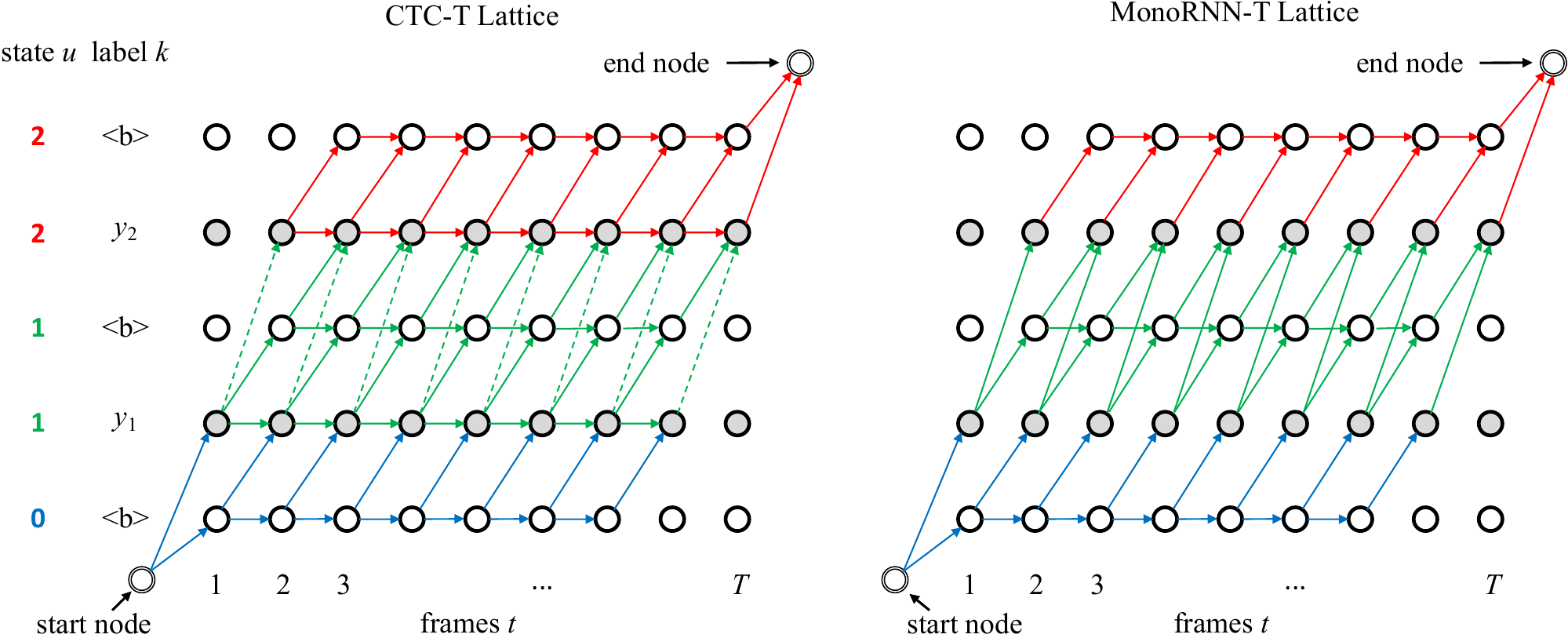}} 
  \caption{ GTC-T lattices for the CTC-T and MonoRNN-T graph topologies. Nodes are represented as circles, edges as arrows, and optional edges by dashed arrows. Optional edges are skipped in CTC-T when two consecutive labels are the same, e.g. $y_1 = y_2$.
  All paths that start at the start node and reach the end node correspond to possible alignments. Edge colors indicate the decoder state $u$. The observation probabilities $\upsilon^k_{t,u}$ are associated with edges, which can be identified by $k$ and $t$ of the destination node in the lattice. $<\!\!\text{b}\!\!>$ denotes the blank symbol.  } 
\label{fig:lattices}
\end{figure*}

The transducer model is an end-to-end ASR architecture that is composed of an encoder, a prediction network, and a joiner network. Its task is to generate a label sequence $Y=(y_1,\dots,y_L)$ of length $L$, e.g., a sequence of words or word-pieces, from an input sequence $X=(\bm x_1,\dots,x_N)$ of length $N$, commonly a sequence of acoustic features such as Mel-spectral energies. The {\em encoder neural network} processes the input sequence $X$ to produce a sequence of acoustic representations $H^\text{enc}=(\bm h^\text{enc}_1,\dots,\bm h^\text{enc}_T)$ of length $T$, which can differ from $N$ due to sub-sampling.
The {\em prediction neural network} acts as an internal language model or decoder to produce a representation $\bm h^\text{dec}_u$, where $u$ denotes the decoder state. Typically, $u$ depends on the previous output labels $y_{0:l-1}=(y_0,\dots,y_{l-1})$,
where $y_0$ corresponds to a start-of-sentence symbol and $l$ to the label index.
Lastly, the {\em joiner network} receives as an input the output representations from the encoder and prediction network to form the joint representation $\bm h^\text{joint}_{t,u}$, where $t$ denotes the encoder frame index.
A probability distribution $\bm \upsilon_{t,u}$ over the set of output labels $\mathcal{U} \cup \{\varnothing\}$, including the {\em blank} symbol $\varnothing$, is derived for each $(t,u)$ point by applying a softmax layer to $\bm h^\text{joint}_{t,u}$.
The observation probability of label $k \in \mathcal{U} \cup \{\varnothing\}$ for the $(t,u)$ point is denoted by $\upsilon^k_{t,u}$.

\subsection{RNN-T}

The RNN-T objective minimizes the loss function $\mathcal{L}_\text{rnnt} = - \ln p(Y|X)$ with
\begin{equation}
\label{eq:rnnt}
   p(Y|X) = \sum_{\pi \in \mathcal{B}^{-1}(Y)} p(\pi | X),
\end{equation}
which corresponds to maximizing the sum of alignment probabilities $p(\pi|X)$ for a set of possible alignments that can be generated from the label sequence $Y$. $\mathcal{B}$ denotes a mapping function that removes all blank symbols from the alignment sequence $\pi$ such that $\mathcal{B}(\pi) = Y$. 
The $p(Y|X)$ for all $1 \leq t \leq T$ and $0 \leq u \leq U$ can be computed efficiently with the forward-backward algorithm. The forward and backward variables, $\alpha$ and $\beta$, are computed by
\begin{equation}
\label{eq:rnnt_alpha_beta}
   \alpha(t,u) = \alpha(t-1,u) \varnothing(t-1,u) + \alpha(t,u-1)y(t,u-1),
\end{equation} 
and
\begin{equation}
   \beta(t,u) = \beta(t+1,u) \varnothing(t,u) + \beta(t,u+1)y(t,u),
\end{equation}
with the initial conditions $\alpha(1,0)=1$ and $\beta(T,U) = \varnothing(T,U)$, where $\varnothing(t,u)$ and $y(t,u)$ denote the blank and label probability for the $(t,u)$ point, respectively.

From this definition follows that RNN-T has a ```square-like'' lattice, since it is allowing label emissions without an advancement in time \cite{Graves12}. The forward and backward variables also allow us to compute $p(Y|X) = \alpha(t,u) \beta(t,u)$ for any $(t,u)$ point. Gradients are derived by computing the derivative of $\mathcal{L}_\text{rnnt}$ with respect to the joiner network output $\bm h^\text{joint}_{t,u}(k)$ before the softmax is applied \cite{Graves12}.

\subsection{Alignment restricted RNN-T}

The {\em alignment restricted} RNN-T (AR-RNN-T) loss utilizes prior alignment information, e.g., forced alignment information from a traditional hybrid acoustic model, to restrict the set of alignments in (\ref{eq:rnnt}) to a set of valid alignments. An alignment it considered valid when the time position of each label in the alignment path is within a predefined time range, which for example is determined from the forced alignment information. The usage of the AR-RNN-T loss allows pruning the joiner output, which significantly improves memory usage and training speed \cite{JayAlignRest2021}.

\subsection{GTC-Transducer}

The GTC-Transducer (GTC-T) loss function is a generalization of the RNNT- loss, defined by $\mathcal{L}_\text{gtct} = - \ln p(\mathcal{G}|X)$ with
\begin{equation}
\label{eq:gtct}
   p(\mathcal{G}|X) = \sum_{\pi \in \mathcal{S}(\mathcal{G},T)} p(\pi | X),
\end{equation}
where $\mathcal{S}(\mathcal{G},T)$ denotes a search function that expands graph $\mathcal{G}$ to all possible alignment paths $\pi$ of length $T$, which here correspond to a sequence of nodes.
Thus, the difference to RNN-T is that possible alignments are not pre-defined by the loss function, but instead are determined through the graph $\mathcal{G}$.
Let's assume that nodes are topologically sorted in a breadth-first search manner and indexed with $g = 0, \dots, G+1$. $0$ and $G+1$ denote the non-emitting start and end nodes.
In GTC-T, the forward variable $\alpha$ for $1 \leq t \leq T$ and all $1 \leq g \leq G$ is computed as
\begin{align}
\label{eq:gtct_alpha}
    \alpha (t,g) 
    &= \sum_{\substack{\pi \in \mathcal{S}(\mathcal{G},T):\\ \pi_{0:t} \in \mathcal{S}(\mathcal{G}_{0:g},t)}}
    \prod_{\tau=1}^t \upsilon_{\tau ,\Upsilon(\pi_{\tau-1},\pi_\tau)}^{\Lambda(\pi_{\tau})} , \nonumber \\
    &= \sum_{\substack{\pi \in \mathcal{S}(\mathcal{G},T):\\ \pi_{0:t} \in \mathcal{S}(\mathcal{G}_{0:g},t)}} 
    \alpha(t-1,\pi_{t-1}) \upsilon_{t ,\Upsilon(\pi_{t-1},\pi_t)}^{\Lambda(\pi_{t})} .
\end{align}
Here, $\mathcal{G}_{0:g}$ denotes the sub-graph of $\mathcal{G}$ containing all paths from node $0$ to node $g$; $\Lambda(\pi_\tau)$ denotes the output label observed at node $\pi_\tau$ of alignment sequence $\pi$ and time index $\tau$; and $\Upsilon(\pi_{\tau-1},\pi_\tau)$ represents the decoder state $u$ for the transition from node $\pi_{\tau-1}$ to $\pi_\tau$.
To compute $\alpha$, the sum is taken over all possible $\pi$ whose sub-sequence up to time index $t$ can be generated in $t$ steps from the sub-graph $\mathcal{G}_{0:g}$, with the initial condition $\alpha(0,g)=1$ for $g=0$.

Likewise, the backward variable $\beta$ is computed by
\begin{align}
\label{eq:gtct_beta}
    \beta (t,g)
    &= \sum_{\substack{\pi \in \mathcal{S}(\mathcal{G},T):\\ \pi_{t:T+1} \in \mathcal{S}(\mathcal{G}_{g:G+1},T-t+1)}}
    \prod_{\tau=t}^{T-1} \upsilon_{\tau+1 ,\Upsilon(\pi_{\tau},\pi_{\tau+1})}^{\Lambda(\pi_{\tau+1})}, \nonumber \\
    &= \!\!\!\!\!\!\!\!\!\!\!\!\!\!\!
    \sum_{\substack{\pi \in \mathcal{S}(\mathcal{G},T):\\ \pi_{t:T+1} \in \mathcal{S}(\mathcal{G}_{g:G+1},T-t+1)}}
    \!\!\!\!\!\!\!\!\!\!\!\!\!\!\!
    \beta (t+1, \pi_{t+1}) \upsilon_{\tau+1 ,\Upsilon(\pi_{\tau},\pi_{\tau+1})}^{\Lambda(\pi_{\tau+1})},
\end{align}
where $\mathcal{G}_{g:G+1}$ denotes the sub-graph of $\mathcal{G}$ containing all paths from node $g$ to node $G+1$.
$\beta$ is initialized with $\beta(T+1,g)=1$ for $g=G+1$ and with $0$ otherwise.
The probability function $p(\mathcal{G}|X)$ can be computed as
\begin{equation}
\label{eq:gtct_prop_fun}
    p(\mathcal{G}|X) = \sum_{(g,g') \in \mathcal{G}} \alpha(t-1,g) \upsilon_{t ,\Upsilon(g,g')}^{\Lambda(g')} \beta(t,g') ,
\end{equation}
where $(g,g')$ denotes an edge in graph $\mathcal{G}$ with the start and destination nodes $g$ and $g'$.
Differentiation of $\mathcal{L}_\text{gtct}$ with respect to $\bm h^\text{joint}_{t,u}(k)$, the joiner network output for label $k$ and $(t,u)$ before the softmax layer, leads to the neural network gradients \cite{moritz2021sequence}.

In this work, two graph topologies are considered that correspond to a CTC-like transducer (CTC-T) lattice and to the MonoRNN-T lattice. They are visualized in Figure~\ref{fig:lattices}.
One of the main differences is that CTC-T allows label repetitions by using self-transitions for nodes where an ASR token is emitted, which is not the case for MonoRNN-T.
For ASR models that do not include a prediction network such as CTC, it can be shown that that a topology which allows label repetitions has better convergence and leads to better results---we do not present results for this experiment in this work but we did test this in the past.
Therefore, the comparison of different label topologies is of high importance also for transducer models in order to find an optimal solution.

\section{Experimental Setup}

The public LibriSpeech ASR corpus and our in-house training and test sets are described in the datasets section. Training settings and model architectures are described in the model setup section.

\subsection{Datasets}

\textbf{LibriSpeech} is collected from recordings of read English audio books with 960 hours of training data plus two development and two test sets, named clean and other \cite{librispeech}. In this work, however, we're only reporting results for the test-clean and test-other conditions with about 5 hours of data each.

Our \textbf{in-house training data} combines two sources. The first consists of 20K hours of English Facebook video data
that is completely de-identified before transcription.
The second contains 20K hours of manually transcribed de-identified English voice-assistant data with no user-identifiable information (UII). All utterances are morphed when researchers manually access them to further de-identify the speaker. Note that the data is not morphed during training. We further augment the data with speed perturbation, simulated room impulse responses, and background noise, resulting in 83M utterances (145K hours).
For evaluation, we consider the following three in-house test sets:

\textbf{VA1} – 10.2K hand-transcribed de-identified short-form utterances (less than five words on average) in the voice-assistant domain, collected from internal volunteer participants. The participants consist of households that have agreed to have their Portal voice activity reviewed and analyzed.

\textbf{VA2} – 44.2K hand-transcribed de-identified short-form utterances in the voice-assistant domain, collected by a third-party data vendor via Oculus devices.

\textbf{Q\&A} – 5.7K hand-transcribed de-identified medium-length utterances (more than 13 words on average) collected by crowd-sourced workers via mobile devices. The utterances consist of
free-form questions directed toward a voice assistant.

\subsection{Model setup}

Two different ASR frameworks are used: ESPnet \cite{watanabe2018espnet} and an internal toolkit based on fairseq and PyTorch \cite{ott2019fairseq}. ESPnet is employed to evaluate different training strategies using a full-sequence model architecture and optimizer settings similar to the LibriSpeech setup of \cite{moritz2021sequence}, except with 14 Conformer blocks instead of 12. The
fairseq-based framework is used for the streaming and large-scale ASR experiments.

The streaming ASR system of this work uses the ``Emformer'' encoder architecture \cite{Shi2021EmformerEM}. Input to the encoder are 4 stacked 80-dimensional spectral-energy feature vectors, which reduces the frame rate to 40~ms.
The predictor consists of 3 LSTM layers with 512 hidden units and layer normalization. Both the encoder and predictor outputs are projected to 1024 dimensional embeddings that are fed to the joiner network, which consist of a projection layer and a softmax layer for a word-piece vocabulary of size 4096.

In the fairseq-based toolkit, an Adam optimizer with a tri-stage learning-rate scheduler is used. For LibriSpeech, models are trained for 120 epochs with a base learning rate of 0.001, a warmup of 10K iterations, and forced annealing after 60 epochs. Experiments on the large-scale in-house data train for 15 epochs using similar model architecture and training hyper-parameters.

Inference uses a time-synchronous beam search algorithm that is adjusted to the transducer model types. The decoding procedure of the CTC-like transducer (CTC-T) is described in \cite{moritz2021sequence}. The MonoRNN-T beam search is based on a modification of this algorithm.
RNN-T decoding is similar to the time-synchronous beam search algorithm described in \cite{SaonTA20,BoyerESPNET21}. 
In all experiments, the beam size is 10.

\section{Training Strategies}
\label{sec:train_strategy}

\begin{table}[tb]
  \caption{ Word error rate [\%] comparison of different training strategies for the CTC-like Transducer (CTC-T) using the LibriSpeech ASR benchmark and an offline Conformer-based model architecture. }
  \label{tab:train_strategies_ctc_t}
  \centering

  \resizebox{.9\linewidth}{!}
  {
  \begin{tabular}{lcccc}
  \toprule
  Training Strategy & \multicolumn{2}{c}{dev} & \multicolumn{2}{c}{test} \\
   & clean & other & clean & other \\
  \hline
  baseline & 3.2 & 8.3 & 3.4 & 8.5 \\
  joint training w/ CTC & 2.9 & 7.6 & 3.1 & 7.6 \\
  joint training w/ LAS & 2.4 & 6.6 & 2.8 & 6.8 \\
  init encoder from CTC & 2.7 & 7.2 & 2.9 & 7.2 \\
  init from RNN-T & 2.5 & 6.7 & 2.7 & 6.7 \\

\bottomrule
  \end{tabular}}
\end{table}

\begin{table}[tb]
  \caption{ LibriSpeech test set word error rates [\%] using an Emformer-based model architecture for streaming ASR. Speed perturbation is applied for data augmentation. }
  \label{tab:results_fairseq_libri}
  \centering
  \resizebox{.99\linewidth}{!}
  {
  \begin{tabular}{lcccc}
  \toprule
  System & \multicolumn{2}{c}{w/o LM} & \multicolumn{2}{c}{w/ LM} \\
   & clean & other & clean & other \\
  \hline
  RNN-T  & 3.25 & 7.97 & 2.86 & 6.90 \\
  RNN-T + init from RNN-T  & 3.49 & 8.39 & 2.98 & 7.32 \\
  AR-RNN-T  & \textbf{3.15} & 7.67 & 2.77 & 6.74 \\
  AR-RNN-T + init from AR-RNN-T  & 3.27 & 7.96 & 2.87 & 6.90 \\
  CTC-T & 3.41 & 8.39 & 2.77 & 6.89 \\
  CTC-T + init from AR-RNN-T  & 3.19 & \textbf{7.66} & \textbf{2.69} & \textbf{6.53} \\
  MonoRNN-T & 3.65 & 8.77 & 2.90 & 7.05 \\
  MonoRNN-T + init from AR-RNN-T  & 3.21 & 7.78 & 2.76 & 6.63 \\

\bottomrule
  \end{tabular}}
\end{table}

Prior work indicated that monotonic transducer losses may require more careful training to reach good performance \cite{moritz2021sequence}. For example, encoder parameters were initialized by CTC pre-training in order to obtain robust results. Here we are comparing different training strategies such as pre-training with CTC or RNN-T, and joint training with an LAS decoder \cite{LAS_Chan16} or CTC.
Note that joint training with an LAS decoder or CTC is solely for regularizing the training process---we did not use the auxiliary outputs for joint scoring purposes.

Table~\ref{tab:train_strategies_ctc_t} presents a comparison of the proposed training strategies using the CTC-T loss function and a Conformer-based offline ASR system implemented in ESPnet \cite{watanabe2018espnet}. The {\em baseline} results are obtained by training the model from scratch and without using any joint training strategy.
It is demonstrated that all four training strategies lead to an substantial WER improvement.
Results show that LAS-based joint training is more effective than CTC-based joined training, where the latter is also a popular training strategy for RNN-T models \cite{BoyerESPNET21}.
Another popular training strategy for transducer models is to initialize encoder parameters by CTC pre-training \cite{Hu2020ExploringPW,Rao2017ExploringAD}. Here it is compared to RNN-T pre-training, which also initializes the prediction network and the joint network. Results show that RNN-T pre-training is more effective than CTC pre-training, with results similar to the aforementioned joint training with an auxiliary LAS decoder.
The two most effective training strategies are joint training with an LAS decoder and parameter initialization from RNN-T pre-training, which both contribute to achieve about similar WERs.
In the further reading we will settle on parameter initialization from RNN-T pre-training as the default training strategy, because this method is easier to implement and does not require any additional model parameters that must be trained.

\section{Comparing Transducer Models}
\label{sec:extra_train}

ASR results for the RNN-T, {\em alignment-restricted} RNN-T (AR-RNN-T), CTC-T, and MonoRNN-T models with and without RNN-T pre-training are shown in Table~\ref{tab:results_fairseq_libri}. For the results presented here, an Emformer-based streaming ASR model architecture is used that is implemented using an internal fairseq-based ASR framework \cite{Shi2021EmformerEM,ott2019fairseq}. The center block size and the look-ahead of the Emformer encoder neural network amount to 1.28s and 240ms, respectively \cite{Shi2021EmformerEM}.

Earlier results discussed in Section~\ref{sec:train_strategy} demonstrated the effectiveness of various training-enhancement strategies for the CTC-T ASR model, where we found that RNN-T parameter initialization followed by 10 epochs of fine-tuning leads to promising results.
However, it is unclear whether the 10 epochs of extra training is part of the reason for the improved WERs, and whether the same additional training would also lead to improved RNN-T results.
Therefore, in Table~\ref{tab:results_fairseq_libri} results are presented for initializing extra RNN-T training from RNN-T pre-training, which results in deteriorated WERs.
For this extra training, the Adam optimizer uses special learning-rate scheduling, where the learning rate is quickly warmed up for 400 iterations and decayed over 10 epochs of extra training using cosine annealing. This is the same setup as is used for CTC-T and MonoRNN-T in this work.
We can conclude that the extra training does {\em not} explain the improved results for CTC-T and MonoRNN-T when using parameter initializing from RNN-T/AR-RNN-T, which can be seen from the results in Table~\ref{tab:results_fairseq_libri}.
Note that we also tested a few different optimizer settings for this experiments, which did not change the fact that this pre-training was not beneficial for RNN-T training.

Table~\ref{tab:results_fairseq_libri} also shows results for the AR-RNN-T loss \cite{JayAlignRest2021}, which outperforms the RNN-T baseline.
However, both monotonic transducer ASR models---CTC-T and MonoRNN-T---obtain similar or better ASR results than the AR-RNN-T when model parameters are pre-trained using AR-RNN-T.
Without pre-training, the CTC-T system also obtains similar or improved WERs compared to RNN-T when using an external LM for inference. However, the RNN-T model achieves lower WERs when no LM is used for decoding, which shows that CTC-T may be more susceptible to external LM information as it was observed in \cite{moritz2021sequence}.

The best ASR results for LibriSpeech are obtained by the CTC-T system with WERs of 2.69\% and 6.53\%---also outperforming MonoRNN-T for all training and test conditions. Note that the presented LibriSpeech WERs correspond to state-of-the-art results for a streaming transducer-based ASR system \cite{Fastemit,Moritz_DCN,MoritzHLR20,Shi2021EmformerEM,Zhang2020TransformerTA}.

\section{Large-scale ASR Results}

\def\d{\hphantom{0}}

\begin{table}[tb]
  \caption{Word error rates [\%] for the large-scale in-house task. The use of an external LM via shallow fusion is indicated by ``ext. LM''. }
  \label{tab:results_large}
  \centering
  \resizebox{.99\linewidth}{!}
  {
  \begin{tabular}{lcccc}
  \toprule
  Test condition & ext. & \multicolumn{3}{c}{System type} \\
    & LM & AR-RNN-T & CTC-T & MonoRNN-T \\
  \hline
  VA1 & \ding{55} & {\d}4.99 & {\d}4.42 & {\d}4.53 \\ 
  VA2 & \ding{55} & 12.01 & 12.09 & 12.03 \\ 
  VA2 & \ding{51} & 10.18 & 10.15 & 10.09 \\ 
  Q\&A & \ding{55} & {\d}6.89 & {\d}6.91 & {\d}6.96 \\ 
  
\bottomrule
  \end{tabular}}
\end{table}

Lastly, Table~\ref{tab:results_large} shows ASR results for our in-house large-scale data set with about 145K hours of training data. Here, the alignment restricted RNN-T (AR-RNN-T) loss, which is has been shown to perform better than RNN-T, is compared against both monotonic transducer loss types. Furthermore, the MonoRNN-T and CTC-T models are derived by one epoch of extra training similar to Section~\ref{sec:extra_train} (but not the AR-RNN-T, for which it does not help per Table~\ref{tab:results_fairseq_libri}).

For the VA1 test condition, which are short-form utterances in the voice-assistant domain collected through Portal, both CTC-T and MonoRNN-T achieve an absolute WER reduction of 0.57\% and 0.46\%, respectively.
For the VA2 and Q\&A test conditions, which are short-form and medium-length voice-assistant queries collected through Oculus and mobile phones respectively, AR-RNN-T, CTC-T, and MonoRNN-T obtain about similar ASR results.
The use of an external language model (LM) is also evaluated for the VA2 test condition. All three transducer models benefit about equally from shallow fusion.
This means that for the large-scale data set we cannot confirm earlier findings of an improved language model susceptibility of monotonic transducer models, as shown in Table~\ref{tab:results_fairseq_libri} and in \cite{moritz2021sequence}.

\section{Discussion}
\label{sec:discussion}

In Section~\ref{sec:extra_train} as well as in \cite{moritz2021sequence} it is demonstrated that monotonic transducer models may be more susceptible to external LM information. The same is true for CTC models that typically show much larger relative improvements compared to RNN-T when using an LM via shallow fusion.
One possible explanation for this observation is that monotonic transducers are less dependent on the prediction network as compared to RNN-T and the prediction network is often identified as a culprit for this behaviour \cite{Variani2020HybridAT,ILMsub2021}.
For example, the CTC-like alignments of the CTC-T model can be achieved even without the presence of the prediction network. In fact, if the prediction network output in CTC-T would be constant, e.g., equal to zero, we would exactly derive CTC. This is not the case for RNN-T.
An RNN-T alignment allows multiple predictions per time step that are obtained by changing the state of the prediction network, which cannot be achieved when the prediction network would be removed or set to a constant output.
This observation could possibly also explain why monotonic transducer models are more difficult to train to achieve optimal results, because the prediction network is in principle not required for the ASR task but can be exploited by the model to further improve results.
This relationship may be harder to learn, which is why the presented training strategies help to further improve the results.

\section{Conclusions}

This paper presents a detailed investigation of monotonic transducer models using the MonoRNN-T loss as well as a transducer loss with a CTC-like topology, named CTC-T. Both were compared to RNN-T and its sibling, the alignment restricted RNN-T (AR-RNN-T) loss.
Experiments were conducted on LibriSpeech as well as on a large-scale in-house data set with about 145K hours of training data.

We demonstrate that advanced training strategies are important for obtaining robust and competitive results with monotonic transducers.
We conjecture that optimal model parameters are more difficult to learn with strictly monotonic outputs since the prediction network can also be neglected, which is not the case for RNN-T.
RNN-T and CTC-based pre-training were compared, where RNN-T based parameter initialization was found to be more effective.
In addition, joint training with an auxiliary CTC or LAS decoder was investigated,
where the latter was found to be more effective.
RNN-T pre-training and joint training with an auxiliary LAS decoder are found to lead to similar ASR results.
Note that the auxiliary decoder is only used to regularize training---we did not evaluate joint scoring during inference.

Overall, ASR results demonstrate that monotonic transducer models, such as CTC-T or MonoRNN-T, perform as well as or better than RNN-T-based models thanks to the training strategies. This is contrary to prior studies where MonoRNN-T was generally found to perform somewhat worse.

State-of-the-art transducer model results for streaming ASR are reported for the LibriSpeech data set, and we provide evidence for the usability of monotonic transducers for ASR systems at production scale. This opens the door to more simplified FST-based decoders.
An analysis of improved emission delays for streaming ASR remains to future work.

\balance
\bibliographystyle{IEEEbib}
\bibliography{refs}

\end{document}